\documentclass[12pt]{article}%
\usepackage{amssymb}
\usepackage{epsfig}

\catcode`\@=11
\def\numberbysection{\@addtoreset{equation}{section}
        \def\theequation{\thesection.\arabic{equation}}}
\def\beq{\begin{equation}}
\def\eeq{\end{equation}}
\numberbysection
\begin{document}
\begin{titlepage}
\begin{center}
\hfill DFF  1/2/01 \\
\vskip 1.in
{\Large \bf Projectors for the fuzzy sphere}
\vskip 0.5in
P. Valtancoli
\\[.2in]
{\em Dipartimento di Fisica dell' Universita', Firenze \\
and INFN, Sezione di Firenze (Italy)}
\end{center}
\vskip .5in
\begin{abstract}
All fiber bundle with a given set of characteristic classes are
viewable as particular projections of a more general bundle called a
universal classifying space. This notion of projector valued field, a
global definition of connections and gauge fields, can be useful to
define vector bundles for non commutative base spaces. In this paper
we derive the projector valued field for the fuzzy sphere, defining
non-commutative n-monopole configurations, and check that in the
classical limit, using the machinery  of non-commutative geometry, the
corresponding topological charges ( Chern class ) are integers.
\end{abstract}
\medskip
\end{titlepage}
\pagenumbering{arabic}
\section{Introduction}

In classical gauge theory, the basic variables like connections and
gauge fields are usually defined locally in space. There is however a
way to descrive gauge theory by globally defined fields, using the
fiber bundle formulation. These globally defined fields are called
projectors and serve to give an alternative description to the
connections which is suitable for generalizations like non-commutative
geometry \cite{1}-\cite{2}. The theorem  of Narasimham and Ramanan
shows that all fiber bundles with a given set of characteristic
classes are viewable as particular projections of a more general
bundle called a universal space bundle. Consider a fiber bundle
determined by the unitary groups $U(p)$ as gauge groups and by basic
space a (compcat ) manifold denoted by $M$. A principal $U(p)$ bundle
$E$ over $M$ with a connection form $\omega$ can be associated to a
projector valued field $P(x) ( x\in M )$. In more general terms, the
space $\Gamma(M,E)$ of smooth sections of  the vector bundle
$E\rightarrow M$ over a compact manifold M is a finite projective
module over the commutative algebra $C(M)$ and every finite projective
$C(M)$-module can be realized as the module of sections of some vector
bundle over $M$. The correspondence is quite general and well known in
the mathematical literature, less known between the physicist
community.

This equivalence has received particular attention in generalizing the
classical geometry to non commutative geometry, and is the nowadays
basis to generalize the concept of vector bundles to the
noncommutative case \cite{3}-\cite{4}. In ref. \cite{5} a
finite-projective-module description of all monopoles configurations
on the 2-dimensional sphere $S^2$ has been presented. In this paper we
outline its generalization to one of the simplest example of
non-commutative manifolds, the fuzzy sphere \cite{a1}-\cite{a2}. We
find that it is simple to give the description of the non commutative
n-monopoles over the fuzzy sphere in terms of $(|n|+1) \ {\rm x} \ (
|n|+1 )$ matrices, having as entries the elements of the basic non
commutative coordinate algebra of the fuzzy sphere.

An alternative, quite involved, formulation of the projectors has been
given in reference \cite{6}, and resembles the classical description
of the projectors over $S^2$ of ref. \cite{7}. Our formulation is
surely original, more easy to deal with and should have all the good
properties of the analogous one over $S^2$ , proposed by Landi.

As an application of our $n-monopoles$ projectors, we show how to
compute their topological charge ( the  Chern class ) in the classical
limit, using the machinery of non commutative geometry, i.e. the
classical Dirac operator introduced in \cite{a3}.

\section{Fuzzy sphere}

The fuzzy sphere is defined through an algebra $M_n$ of $n \ {\rm x }
\ n$ complex matrices. When $n \rightarrow \infty$ the corresponding
matrix geometry tends to the geometry of the 2-dimesional sphere. In
order to realize such an idea, one views the ordinary geometry of the
sphere as a commutative algebra. A process of truncations of it
produces the matrix algebras, non-commutative algebras which
approximate the commutative algebra in the large $n$ limit.

The $S^2$ sphere can be defined as the constraint between the
coordinates:

\beq g_{ij} x^i x^j = r^2 \label{1} \eeq

The geometry of $S^2$ can be formulated in terms of the algebra ${\cal
C} ( S^2)$ of functions $f(x^i)$ on $S^2$ admitting a polynomial
expansion in the $x^i$:

\beq f(x^i) = f_0 + f_i x^i + \frac{1}{2} f_{ij} x^i x^j + .... \label{2}
\eeq

Let us truncate this algebra by requiring that all coefficients in the
expansion apart from $f_0$ and $f_i$ are zero. In order to turn this
linear approximation to a true algebra, it is possible to replace the
$x^i$ with the algebra $M_2$ of complex $2 \ {\rm x} \ 2$ complex
matrices. Therefore we identify $x^i$ with the algebra of Pauli
matrices:

\beq x^i \rightarrow x^i = \frac{\alpha}{2} \sigma^i \label{3} \eeq

Moreover the Casimir of the algebra can be taken as a constant and
coincides with the constraint (\ref{1}) , by imposing that

\beq \alpha^2 = \frac{4}{3} r^2 \label{4} \eeq

The algebra $M_2$ describes a fuzzy sphere, where only the north and
south poles can be distinguished.

Keeping the term quadratic in the $x^i$, the quadratic approximation
to the commutative algebra of functions can be turned into an algebra,
by identifying $x^i$ with the algebra $M_3$ of complex $3 \ {\rm  x }
\ 3$ matrices. Generalizing this truncation to the $n$-term of the
series (\ref{2})  we can define the general fuzzy sphere in terms  of
the algebra $M_n$ of complex $n \ {\rm  x} \ n$ matrices, through the
identification

\beq x^i \rightarrow x^i = \alpha J^i \label{5} \eeq

where $J^i$ is a $n \ {\rm x} \ n$ representation of the $SU(2)$
algebra.

Therefore the fuzzy sphere
\cite{a1}-\cite{a2}-\cite{a3}-\cite{a4}-\cite{a5}-\cite{a6} can be
defined by the commutation relations:

\begin{eqnarray} [x_i, x_j ] & = & i \alpha \epsilon_{ijk} x_k \nonumber \\
\alpha & = & \frac{r}{\sqrt{j(j+1)}} \nonumber \\
\sum_i {(x^i)}^2 & = & r^2 \label{6} \end{eqnarray}

This can be realized as an algebra of a couple of oscillators

\begin{eqnarray}
x_1 & = & \frac{\hat{\alpha}}{2}( a_0 a^{\dagger}_1 + a_1
a^{\dagger}_0 ) \nonumber
\\ x_2 & = & \frac{\hat{\alpha}}{2} i ( a_0 a^{\dagger}_1
- a_1 a^{\dagger}_0 ) \nonumber \\   x_3 & = &
\frac{\hat{\alpha}}{2} ( a^{\dagger}_0 a_0 - a^{\dagger}_1 a_1 ) \nonumber
\\ \hat{N} & = & a^{\dagger}_0 a_0 + a^{\dagger}_1 a_1 \label{7}
\end{eqnarray}

together with the commutation relations

\beq [ a_i , a^{\dagger}_j ] = \delta_{ij} \label{8} \eeq

where the operator $\hat{\alpha}$ is defined as

\beq \hat{\alpha} = 2\frac{r}{\sqrt{\hat{N} ( \hat{N}+2 )}}  \label{9} \eeq

and is equal to $\alpha$ for representations at fixed number $\hat{N}
= N = 2j $, as we demand to build our fixed-$j$ fuzzy sphere.

In the classical limit $N\rightarrow \infty$ this construction
corresponds to the $U(1)$ principal (Hopf) fibration $\pi : S^3
\rightarrow S^2 $ over the two dimensional sphere $S^2$ \cite{5}.
An ambiguity of a $U(1)$ factor into the definition of the oscillators
is cancelled in the $x^i$ combinations. The same trick is useful to
compute the projectors in terms of vector valued fields. They carry
the $U(1)$ ambiguity, which is cancelled in the combination of bra and
ket vectors giving rise to the projectors. Therefore our vector fields
will be defined in terms of the oscillators while the projectors will
be functions only of the $x_i$ algebra.

\section{The projectors for n-monopoles}

The monopole connections can be alternatively described by some
projected valued fields $P_n (x)$. The canonical connection associated
with the projector $P_n (x)$ has curvature given by

\beq \nabla^2 = P_n dP_n dP_n \label{10} \eeq

To construct the n-monopole projectors $P_n(x)$ for the fuzzy sphere
consider the n-dimensional vectors
\beq | \psi_{n} > = N_n \left( \begin{array}{c} {(a_0)}^n \\
... \\ \sqrt{\left( \begin{array}{c} n \\ k \end{array} \right)}
{(a_0)}^{n-k} {(a_1)}^k \\ ... \\ {(a_1)}^n \end{array} \right)
\label{11} \eeq

The normalization condition for these vectors fixes the function $N_n$
to be dependent on the number operator $\hat{N} = a^{\dagger}_0 a_0 +
a^{\dagger}_1 a_1$:

\begin{eqnarray} < \psi_{n} | \psi_{n} > & = & 1 \nonumber \\
 N_n = N_n ( \hat{N} ) & = & \frac{1}{\sqrt{ \prod_{i=0}^{n-1} ( \hat{N}
- i  + n ) }} \label{12} \end{eqnarray}

The corresponding $n$-monopole connection $1$-form has a very simple
expression in terms of the vector valued function $|\psi_n>$ :

\beq A^\nabla_n = <\psi_n | d \psi_n > \label{13} \eeq

The projector for the  $n$-monopole is defined to be:

\beq P_n = |\psi_{n} > < \psi_{n} | \ \ \ \ \ \ n < N+2 \label{14} \eeq

and satisfies the basic properties of a projector, being

\beq P_n^2 = P_n \ \ \  \ \  P^{\dagger}_n = P_n \label{15} \eeq

It is easy to notice that in the  product there appears only
combinations of oscillators that can be written in terms of the
algebra $x_i$ of the fuzzy sphere. Therefore $P_n$ is a matrix having
as entries the basic operator algebra of the theory.

Consistently the projector $P_n$ has a nonvanishing positive trace
given by:

\begin{eqnarray} Tr P_n & = & Tr |\psi_{n} > < \psi_{n} | =
N_n \prod^{n-1}_{i=0} ( a^{\dagger}_0 a_0 + a^{\dagger}_1 a_1  + i +
2) N_n
= \nonumber \\ & = & \frac{ \hat{N} + n + 1}{\hat{N} + 1}  \label{16}
\end{eqnarray}

In the classical limit $N \rightarrow \infty$ the trace is equal 1,
and the corresponding classical projectors \cite{5} are matrices of
numbers of rank $1$.

\section{The projectors for n-antimonopoles}

To construct the n-antimonopole solution it is enough to take the
adjoint of the vectors (\ref{11}). Consider the n-dimensional vectors
\beq | \psi_{-n} > = N_n \left( \begin{array}{c} {(a^{\dagger}_0)}^n \\
... \\ \sqrt{ \left( \begin{array}{c} n \\ k \end{array} \right) }
{(a^{\dagger}_0)}^{n-k} {(a^{\dagger}_1)}^k \\ ... \\
{(a^{\dagger}_1)}^n \end{array} \right)
\label{17} \eeq

The normalization condition for these vectors fixes the function $N_n$
to be dependent on the number operator $\hat{N} = a^{\dagger}_0 a_0 +
a^{\dagger}_1 a_1$:

\begin{eqnarray} < \psi_{-n} | \psi_{-n} > & = & 1 \nonumber \\
 N_n = N_n ( \hat{N} ) & = & \frac{1}{\sqrt{ \prod_{i=0}^{n-1} ( \hat{N}
+ i + 2 - n ) }} \label{18} \end{eqnarray}

The corresponding projector for antimonopoles is:

\beq P_{-n} = |\psi_{-n} > < \psi_{-n} | \ \ \ \ \ \ n < N+2 \label{19} \eeq

Consistently the projector $P_{-n}$ has a trace given by:

\begin{eqnarray} Tr P_{-n} & = & Tr |\psi_{-n} > < \psi_{-n} | =
N_n \prod^{n-1}_{i=0} ( a^{\dagger}_0 a_0 + a^{\dagger}_1 a_1 - i )
N_n = \nonumber \\ & = & \frac{ \hat{N} +1 - n }{\hat{N}+1}
\label{20} \end{eqnarray}

This trace is positive definite if and only if the following bound is
respected:

\beq n < N+1 \label{21} \eeq

\section{Classical limit and Chern Class}

The projectors above, defining the $n$-projective moduli of the
non-commutative theory, can be checked to reproduce the known results
for the classical sphere ( see Landi \cite{5} ). We use here the
tecnology of the non-commutative geometry to compute the first Chern
class, and therefore

\beq c_n = \frac{1}{2\pi} \int d\Omega \ Tr \ ( \gamma_5  P_n d P_n d P_n )
\label{22} \eeq

It is not difficult to manipolate the Chern class

\beq P_n d P_n d P_n = |\psi_n> \ \{  < d\psi_n | d \psi_n > +
{( < \psi_n |  d \psi_n > )}^2 \} \ <\psi_n | \label{23} \eeq

We use as derivation the commutator of the classical Dirac operator on
the sphere \cite{a3}:

\beq D = \sigma^i \cdot L^i + 1 \label{24} \eeq

that anticommutes with the classical $\gamma_5$ operator \cite{a4}
given by

\beq \gamma_5 = \sigma^i \cdot \frac{x^i}{r} \label{25} \eeq

The Lie derivative $L_i$ has as commutation relations

\beq [L^i, L^j] = i \epsilon_{ijk} L^k \label{26} \eeq

and can be taken as the following adjoint action:

\beq L^i = [ \frac{x^i}{\hat{\alpha}}, . ] \label{27} \eeq

In order to compute the n-monopole 1-form connection, we first compute

\begin{eqnarray} < \psi_n | \frac{x_i}{\hat{\alpha}} |\psi_n> & = &
\frac{\hat{N}  - n}{\hat{N}  } \frac{x_i}{\hat{\alpha}}\nonumber \\
< \psi_{-n} | \frac{x_i}{\hat{\alpha}} |\psi_{-n}> & = &
\frac{\hat{N} + 2 + n}{\hat{N} + 2 } \frac{x_i}{\hat{\alpha}}
\label{28} \end{eqnarray}

by observing that $x_i$ is always a combination of oscillators $a_i$
and their adjoint $a^{\dagger}_j$. These can be commuted with $
|\psi_{\pm n}> N^{-1}_n ( \hat{N} \mp n )$ and $N^{-1}_n( \hat{N}
\mp n) <\psi_{\pm n}|$ and the resulting scalar product is just the
normalization condition.

Therefore the Lie derivative , applied to the oscillators, gives rise
to:

\begin{eqnarray} <\psi_{n} | \sigma_i L_i |\psi_{n}> & = & - \frac{n}{\hat{N}}
\frac{\sigma^i x^i}{\hat{\alpha}} \nonumber \\
 <\psi_{-n} | \sigma_i L_i |\psi_{-n}> & = &  \frac{n}{\hat{N}+2}
\frac{\sigma^i x^i}{\hat{\alpha}} \label{29} \end{eqnarray}

Since ${(\sigma^i x^i)}^2 = r^2 - \hat{\alpha}  \ \sigma^i x^i$ we can
compute the second term in eq. (\ref{23}):

\begin{eqnarray} {( < \psi_n |  d \psi_n > )}^2 & = & \frac{n^2}{\hat{N}^2}
( \frac{r^2}{\hat{\alpha}^2} - \frac{\sigma^i x^i}{\hat{\alpha}} )
\nonumber \\
 {( < \psi_{-n} |  d \psi_{-n} > )}^2 & = & \frac{n^2}{{(\hat{N} + 2)}^2 }
( \frac{r^2}{\hat{\alpha}^2} - \frac{\sigma^i x^i}{\hat{\alpha}} )
\label{30} \end{eqnarray}

Its contribution to the first Chern class is vanishing in the
classical limit, since the only surviving term to the trace is
depressed by a factor $\frac{1}{N}$.

Let us compute the first term of eq. (\ref{23}) which should give the
real contribution:

\begin{eqnarray} < d\psi_n | d\psi_n > & = & \nonumber \\
& = & \frac{\sigma^i x^i}{\hat{\alpha}} <\psi_n |
\frac{\sigma^i x^i}{\hat{\alpha}} |\psi_n > -
{(\frac{\sigma^i x^i}{\hat{\alpha}})}^2 - <\psi_n| {(\frac{\sigma^i
x^i}{\hat{\alpha}})}^2 |\psi_n > + <\psi_n |\frac{\sigma^i
x^i}{\hat{\alpha}}|\psi_n > \frac{\sigma^i x^i}{\hat{\alpha}}
\nonumber \\ & = &
n \frac{\sigma^i x^i}{ N \alpha} - \frac{n ( 2 + n )}{4}
\label{31} \end{eqnarray}

and

\beq < d\psi_{-n} | d\psi_{-n} > = - n
\frac{\sigma^i x^i}{ (N + 2 ) \alpha} - \frac{n (  2 + n )}{4}
\label{32} \eeq

In the classical limit the first term in r.h.s. of eqs. (\ref{31}) and
(\ref{32}) is equivalent to $ n \gamma^5 / 2 $ while the other term is
cancelled by the trace. Therefore the total contribution is given by:

\beq c_n = \frac{1}{2\pi} \int d\Omega \ Tr ( \gamma^5 \ P_n dP_n dP_n ) \ = \
n (\gamma^5)^2 \ = \ n \label{33} \eeq

\section{Conclusion}

An alternative description to the connections and curvature on the
fiber bundle can be obtained immerging the fiber bundle at fixed
toplogy into a trivial universal bundle, from whom it can be recovered
with a projection. The projected valued field give a complete global
description of the vector bundle, within each topology sector.

 In this paper we have shown how to derive the exact expressions for
the non commutative $n$-monopole on the fuzzy sphere. We have checked
that their topological charge ( first Chern class for the vector
bundle ) are integers in the classical limit, using the classical
expression of the Dirac operator. It would be interesting to verify
that these topological charge are integers in the full non-commutative
theory, but this requires the knowlegde of the full non-commutative
Dirac operator on the sphere. A proposal of it is given in
\cite{a5}-\cite{a6}. However, to our knowledge, it remains to be
checked that it verifies all axioms of non-commutative geometry.

\end{document}